\documentclass[prd,aps,superscriptaddress,twocolumn,floatfix,nofootinbib]{revtex4-1}
\pdfoutput=1

\usepackage{amsfonts}
\usepackage{amsmath}
\usepackage{amssymb}
\usepackage{bm}
\usepackage{dcolumn}
\usepackage{graphicx}   
\usepackage[latin1]{inputenc}
\usepackage{latexsym}
\usepackage{rotating}
\usepackage{hyperref}
\usepackage{graphicx}
\usepackage{color}

\newcommand\be{\begin{equation}}
\newcommand\ba{\begin{eqnarray}}
\newcommand\ee{\end{equation}}
\newcommand\ea{\end{eqnarray}}

\begin{document}

\title{%History of Inflationary Preheating - A Personal Account 
Inflationary Reheating to Preheating - A Personal Account }

\author{Robert Brandenberger}
\email{rhb@physics.mcgill.ca}
\affiliation{Department of Physics, McGill University, Montr\'{e}al,
  QC, H3A 2T8, Canada}
 
 \author{Jennie Traschen}
 \email{traschen@umass.edu}
 \affiliation{Department of Physics, University of Massachusetts, Amherst, MA 01003, USA}
 
%%%%%%%%%%%%%%%%%%%%%%%%%%%%%%%%%%%%%%%%%%

\begin{abstract}

This is a personal account of the early work that led to what is now known as the {\it preheating stage} of inflationary cosmology. The broader applicability of the underlying instability mechanisms in cosmology are indicated.

\end{abstract}

\maketitle

%%%%%%%%%%%%%%%%%%%%%%%%%%%%%%%%%%%%%%%%%%
 
\section{Introduction} 
\label{sec:intro}

In 1990 the authors posted the paper ``Particle Production During Out-of-Equilibrium Phase Transitions", \cite{TB} a fine title in its way, but obscure as to its focus on populating the universe with ordinary matter, following a cosmological constant dominated phase.  By July 1994 the title of our publication, with co-author Yuri Shtanov, had evolved to the more pointed ``Universe Reheating after Inflation" \cite{STB}, whose posting closely followed that of Kofman, Linde, and Starobinsky in May 1994, entitled ``Reheating after Inflation" \cite{KLS1}.  By 1997 this latter set of authors, along with co-author Greene, had evolved the title of related work to ``Structure of Resonance in Preheating after Inflation" \cite{KLS3}, and ``preheating" has perhaps  been the favored term since then.  

The inflationary scenario \cite{Early, Guth} has become the paradigm of early universe cosmology \footnote{It is important, however, to keep in mind that there are alternative scenarios (see e.g. \cite{alternatives} for a discussion of some alternatives),  and that  concerns have been raised about the viability of any effective field theory description of inflation (see e.g. \cite{swamp, TCC}).}.  Cosmological inflation is a period in the evolution of the early universe when the size of 
the observable universe
 expands almost exponentially, thus rendering space to be approximately flat, providing an explanation for the near isotropy of the cosmic microwave background. Further, it gives a causal mechanism for generating cosmological fluctuations \cite{ChibMukh} and gravitational waves \cite{Starob}.  

The standard way of obtaining inflation was to assume that there is new scalar field $\phi$ (the ``inflaton'') having a very flat potential $V(\phi)$ such that its energy-momentum tensor is dominated by the almost constant potential energy contribution \cite{Guth}. At the end of the period of accelerated expansion, the universe is effectively empty of regular matter, and thus an essential part of the inflationary scenario is the presence of a {\it period of reheating},  or as it came to be called, {\it preheating},  during which the energy of the inflaton field is transferred into the matter and radiation which we observe today.

What follows is a personal account of the initial stages of the development of our understanding of this reheating process.
A brief discussion of other applications of the mathematics which was developed to describe reheating follows.
   
%%%%%%%%%%%%%%%%%%%%%%%%%%%%%%%%%%%%%%%%%%%
\section{Initial Work}
\label{section2}
 
 In the early 1980s, the two of us were PhD students at Harvard, working on constructive quantum field theory and general relativistic constraints on cosmological fluctuations, respectively. Through a ``secret'' graduate student seminar we had gotten interested in early universe cosmology, in particular in the exciting development of the inflationary universe scenario and semi-classical particle production. 
 
 At the time,  {\it new inflation} \cite{Linde, AS} was considered to be the paradigm of inflationary cosmology.  Inflation is triggered by the scalar field $\phi$ slowly rolling down its flat potential.  The period of inflation ends when $\phi$ gets sufficiently close to the minimum of $V(\phi)$. At that point, $\phi$ begins to oscillate about the minimum with an amplitude which is damped by the expansion of space.  The production of particles, or reheating process, was initially studied perturbatively by computing the decay rate $\Gamma$ of an inflaton quantum into matter particles $\chi$ to leading order in perturbation theory (assuming a direct coupling between 
 $\phi$ and $\chi$ in the Lagrangian). The decay rate
 $\Gamma$ was then inserted 
 as an effective damping term in the equation of motion for a homogeneous $\phi$ field in a Friedmann-Lemaitre-Robertson-Walker metric with scale factor $a(t)$ and Hubble expansion parameter $H(t) = {\dot a} / a$, where the derivative is with respect to time $t$
 \be
 {\ddot \phi} + \bigl( 3 H + \Gamma \bigr) {\dot \phi} \, = \, - V^{\prime}(\phi) \, ,
 \ee
 where a prime is the derivative with respect to $\phi$ \cite{Dolgov, Abbott, AA}\footnote{In this context, the term ``reheating" appeared at least as early as 1982,  in \cite{AA}. }.  By this process, the energy is gradually transferred to matter.  Since the coupling between $\phi$ and $\chi$ has to be weak in order for quantum effects not to destroy the flatness of the potential $V(\phi)$,  the value of $\Gamma$ is typically much smaller than the Hubble expansion rate $H$ at the end of the inflationary period, which implied that the energy loss of $\phi$ is mainly due to the expansion of space.  Energy transfer to matter can only become efficient once $H$ drops below the value of $\Gamma$. Thus, the temperature of matter (denoted by $T_R$) once most of the energy in $\phi$ has been transferred to it is low compared to the energy scale at the end of inflation is
 \be
 T_R^4 \, \sim \, \Gamma^2 m_{pl}^2 \, ,
 \ee
 where $m_{pl}$ is the Planck mass.  This is parametrically much smaller than the energy density $\rho_f$ at the end of inflation which is
 \be
 \rho_f \, \sim H_f^2 m_{pl}^2 \, ,
 \ee
where $H_f$ is the value of $H$ at the end of inflation. Thus, in this approach to reheating, the energy transfer is slow and the effective temperature of matter after the energy transfer is low.
 
However, we realized that this approach was missing crucial physics. First, the one loop perturbative computation misses the fact that the initial state is a coherently oscillating homogeneous field configuration and not a state made up of isolated inflaton quanta. Second, it does not take into account the fluctuation-dissipation theorem the fact that there are always fluctuations whenever there is damping \footnote{See \cite{Ramos} for an approach which takes the fluctuation-dissipation theorem into account in the context of perturbative inflationary reheating).}. Third, there are ways to produce particles faster and in greater number.

From the Ehrenfest Theorem it follows that the dynamics of the large amplitude homogeneous inflaton field configuration $\phi$ can be described classically, whereas the matter field is unexcited at the end of inflation and hence has to be treated quantum mechanically. Thus, we proposed to study the quantum evolution of $\chi$ in the background of a homogeneously oscillating classical $\phi$ field.  Consider a toy model for the interaction of the inflaton field $\phi$ with a scalar matter field $\chi$ with interaction Lagrangian
\be
{\cal{L}}_{{\rm{int}}} \, = \, - \frac{1}{2} g^2 \chi^2 \phi^2 \, ,
\ee
where $g$ is a dimensionless coupling constant. In a FLRW cosmology with line element
\be
ds^2 \, = \,  -dt^2 + a^2 (t)  \delta_{ij} dx^i dx^j 
\ee
and the oscillating background scalar field 
\be\label{phi}
\phi \, = \, M_0 cos(\omega t )
\ee
the equation for the Fourier mode $ \chi_k$ is
\be \label{eq1}
{\ddot{\chi}}_k + 3{\dot{a} \over a} +  \bigl( k^2 + m_{\chi}^2 + g^2 M_0^2 {\rm{cos}}^2 (\omega t) \bigr) \chi_k \, = \, 0 \, ,
%\chi_k^{\prime\prime} + \bigl( k^2  - {a^{\prime\prime}\over a} +  a^2 m_{\chi}^2 + a^2 g^2 cos (\omega t) \bigr) f_k \, = \, 0 \, ,
\ee
where $m_{\chi}$ is the mass of the $\chi$ field, and self-interactions of $\chi$ have been neglected. In the absence of expansion, so that $a=1$, this is the Mathieu equation
which (as is well known from both the classical mechanics \cite{LL, Arnold} and mathematics \cite{Mathieu} literature) admits instabilities  in resonance bands centered at the frequencies $\omega_k =\sqrt{k^2 + m_{\chi}^2 } = n \omega /2$, $n$ an integer. In these bands there are
growing modes  
\be\label{growing}
\chi_k \, \sim \, e^{\mu_k t }
\ee
and working in perturbation theory for small $M_0^2 / \omega^2 $ the $n=1$ band dominates. Setting $\omega_k = \omega /2 + \delta_k$ one finds that for $|\delta_k | < M_0^2 / 2 \omega $ the Floquet index is
\be\label{index}
  \mu_k \, = \, {M_0^2 \over 2 \omega } \left( 1- {2\omega \delta_k \over M_0^2 } \right)
  \ee 
The production of growing $\chi_k$ modes is then translated into particle production. Hence, the instability due to the matching of mode frequencies to the frequency of oscillation of $\phi$, is what will allow the universe to ``reheat"  with non-inflaton particles.

In the expanding universe, the resonance criteria changes with evolution of $a(t )$, so that different $k$-modes redshift into  and out of the resonance band. While a $k$-mode is within the resonance band, the background expansion can be neglected assuming 
the oscillation of the inflaton is fast compared to the Hubble parameter. The role of the expansion is to then shift a new set of modes into  resonance, so that at the end,  one integrates over all the produced particles of different frequencies during the relaxation phase of $\phi$. Initially, we worked out the dynamics in the context of the {\it new inflation} scenario \cite{Linde, AS}
and showed that this parametric resonance decay channel can be effective and lead to energy transfer from $\phi$ to $\chi$ within a Hubble expansion time (which in particular implies that neglecting the expansion of space is a self-consistent approximation). 

On May 9, 1984,
 we jointly presented our work at the Harvard University Gauge Seminar,  at that time the most prestigious seminar series at Harvard.
Consistent with its prestige, the seminar was
 held in a small room in the Lyman Laboratory wing of the Harvard Physics Department, a room with a single couch (reserved for the senior faculty), a small blackboard, and mostly standing room for the rest of the audience. The title of our presentation was ``Reheating in Inflationary Cosmology''. Sidney Coleman was one of the faculty sitting on the couch, and he was always a step ahead of us. We started by writing down the equation (\ref{eq1}) and he immediately spoke up and said ``parametric resonance''. At the end of the seminar we felt that our work was a trivial application of some well known classical mechanics phenomenon, and that it was not worth writing up.

We did not, however, forget our work, and published it in 1990 \cite{TB}.  The core of that paper was to analyze the particle production due to parametric resonance in the expanding universe, sum over the different resonance bands, and compute the energy in produced $\chi$ particles as the inflaton oscillates and settles into the new vacuum. Unfortunately -- perhaps typically for young researchers -- this punch line calculation was presented in section 6 of the paper, following detailed calculations of the much smaller amount of particle production during the slow-roll evolution of $\phi$ \footnote{At around the same time the paper \cite{DK} appeared in which the presence of a parametric resonance instability in the reheating dynamics was also pointed out.}.

 \section{Preheating Revealed}
 
Our work initially had no impact on the field.  It was not cited until 1994 (modulo two citations, one of them a self-citation in a review article).  Neither of us spoke extensively about this work. But in 1990 we had both been selected to be among a small group of North America-based young cosmologists who were able to travel to Moscow to meet up with similar size group of young Soviet cosmologists. During that meeting we discussed our work extensively with our Soviet colleagues, some of whom (Lev Kofman and Alexei Starobinsky) had also been thinking about non-perturbative approaches to reheating.  We emphasized the role of the parametric resonance instability.  We also started a project with Yuri Shtanov (one of the Soviet participants at the meeting who was from Kiev and who had independently been working on reheating \cite{Shtanov}) aimed at extending and improving our work, in light of the realization that {\it chaotic} inflation \cite{chaotic} is a more consistent scenario for inflation than {\it new inflation}. At that time this realization  was well known behind \footnote{From our perspective.} the iron curtain, but not in front of it. We were able to show how to better include the expansion of space in the analysis: by rescaling the matter field via
\be
\chi_k \, \equiv \, a^{-3/2} Y_k 
\ee
we were able to show that a Floquet type equation for $Y_k$ (the equation of a harmonic oscillator with a periodic correction term to the mass) results, and that thus the exponential increase of $\chi_k$ (modulated by the damping $a^{-3/2}$) persists.  For an interaction Lagrangian
\be
{\cal{L}}_{{\rm{int}}} \, = \, \bigl(\sigma \phi + h \phi^2 \bigr) \chi^2 \, ,
\ee
where $\sigma$ and $h$ are coupling constants,
% (and $\sigma$ has the dimension of mass), 
the equation for $Y_k$ is \cite{STB}
\be
{\ddot{Y}}_k + \bigl( \omega_k^2(t) + g(mt) \bigr) Y_k \, = \, 0 \, ,
\ee
with
\ba
\omega_k^2(t) \, &=& \, k^2 + m_{\chi}^2 - \frac{9}{4}H^2 - \frac{3}{2} {\dot{H}} +  h M_0^2 \nonumber \\
g(mt) \, &=& \, 2 \sigma \phi + 2 h \bigl(\phi^2 - \frac{1}{2} M_0^2 \bigr) \, .
\ea
with $\phi$ given in (\ref{phi}).  Note that $g(mt)$ is periodic (with period $2\pi / m$). Floquet theory \cite{Floquet} then tells us that there will be exponential growth of $Y_k$ in specific resonance bands. We worked this out explicitly in \cite{STB},  and also worked out the particle production in a fermionic field.

Our work \cite{STB} was ready for submission in 1994.  Shortly before,  Kofman, Linde and Starobinsky released their short paper \cite{KLS1} in which they invented the term {\it preheating} to describe the intial stage of energy transfer from the inflaton to matter.  Importantly, they focused attention on the broad parametric resonance channel which is more efficient that the narrow band channel.  Suddenly,  cosmologists working on inflation began to pay attention to the resonant instabilities which lead to rapid energy transfer, and our paper began to rapidly accumulate citations. Kofman, Linde and Starobinsky  then worked out preheating in detail in a beautiful paper \cite{KLS2} which led to an improved analytical understanding of the processes, in particular taking the expansion of space into account \footnote{Note also the detailed studies of particle production and scalar field dynamics in \cite{Boy}}.

In the following years there was much research
studying preheating in various inflationary models. In particular, it was discovered that in some cases there is an even more efficient tachyonic resonance \cite{tachyon}.  Preheating into fermions was further studied
 (although it is less efficient because of the Pauli exclusion principle) \cite{Fermions}.  The oscillations of the inflaton field will also lead to graviton production \cite{graviton}. For reviews of the large body of work on preheating see e.g. \cite{ABCM, Karouby, Lozanov}.

 \section{Preheating and Anderson Localization}
 
The preheating instability does not go on forever, but is terminated due to back-reaction effects.  The most basic criterion is that not more energy can be produced in $\chi$ particles than is available in the initial $\phi$ condensate.  The production of $\chi$ particles will in fact reduce the amplitude of the condensate, and, taking this into account, the equation for $\chi$ particles will no longer have an exactly periodic mass term.  Next, as a consequence of the nonlinearity of the system,  fluctuations in $\phi$ will be generated, and they will eventually destroy the coherence of the condensate.  To study these effects, codes were developed which simulate the classical two field dynamics. Some of the codes are {\it LATTICEEASY} \cite{Felder},  {\it DEFROST} \cite{Frolov}, {\it CosmoLattice} \cite{Figueroa},  {\it PSpectRe} \cite{Easther} and {\it Gabe} \cite{Giblin}. Results obtained using these numerical codes demonstrate the robustness of the preheating channel.

However, another worry arose: in dynamical systems there are inevitably sources of noise.  Is the preheating channel stable in the presence of noise? The first
expectation was that the instability could persist but with a slightly weaker strength,  but initial numerical studies indicated that noise could in fact make the resonance stronger.  Walter Craig, at that time a mathematics professor at Brown University, became interested in this question, and he and one of us (RB, at the time at Brown) started collaborating. We added a homogeneous random noise term $n(t)$ to the differential equation (\ref{eq1}) of preheating, obtaining
\be \label{eq3}
{\ddot{\chi}}_k + 3{\dot{a} \over a} +  \bigl( k^2 + m_{\chi}^2 + g^2 M_0^2 {\rm{cos}}^2 (\omega t) + n(t) \bigr) \chi_k \, = \, 0 
%\chi_k^{\prime \prime} + \bigl(\omega^2_k + q {\rm{cos}}(2z) + n(z) \bigr) \chi_k \, = \, 0 \, .
\ee
Under the assumption that the noise term $n(t)$ is uncorrelated in different oscillation periods of the term 
$g^2 M_0^2 {\rm{cos}}^2 (\omega t)$ driving the resonance, and that it is identically distributed, we were able to show \cite{Craig1} that
\be
\mu_k (n \neq 0) \, > \, \mu_k(n = 0) \, ,
\ee
where the left hand side is the value of the Floquet exponent in the presence of noise and the right hand side the corresponding exponent in the absence of noise. The strict inequality sign in the above leads to the surprising result that not only does the noise make the resonance stronger within a resonance band, but, in fact, the noise term eliminates the stability bands - all modes become unstable! 

In \cite{Craig2} we extended the study to the case of inhomogeneous noise, and concluded that for inhomogeneous noise the Floquet exponent of each mode is strictly larger than the maximal Floquet exponent in the absence of noise, assuming certain properties of the noise.

As T.D. Lee (in a seminar given by one of us (RB) at Columbia University) and Walter Craig realized (independently),  the analysis of \cite{Craig1} provides a new proof of Anderson Localization in 1d \cite{Craig3}.  This is based on the well-known correspondence between a classical field theory problem and a one-dimensional quantum mechanics problem. If we replace time $t$ by the spatial coordinate $x$,  and the field $\chi$ by the wave function $\psi$,  our equation (\ref{eq3}) becomes the time-independent Schr\"{o}dinger equation
\be \label{eq4}
H \psi \, = \, E \psi \, ,
\ee
where $E = \omega_k^2$ and the Hamiltonian $H$ is 
\be
H \, = \, \frac{\partial^2}{(\partial x)^2} + V_p(x) + V_R(x) \, ,
\ee
with a periodic contribution $V_p(x)$ to the potential and a random noise term $V_R(x)$. In the absence of noise, the solutions of (\ref{eq4}) are periodic functions, the Bloch wave states. Our theorem implies that in the presence of random noise obeying our conditions, the states $\psi$ become localized, i.e. exponentially decaying (due to normalizability of $\psi$ there is no exponentially growing term).  Thus, the noise destroys the Bloch wave states. This is Anderson localization.

\section{Further Applications}
 
The mathematics of (p)reheating  has many other applications, in particular other applications in cosmology. In many theories beyond the Standard Model there are scalar fields called {\it moduli fields} which are assumed to be coherently oscillating (due to some initial misalignment mechanism such as recently proposed in \cite{misalign}).  Such moduli fields can excite parametric resonance instabilities in all fields which they couple to.  This can induce moduli field decay \cite{Natalia},  or moduli trapping at enhanced symmetry points \cite{Maloney}.

Another application is to ultralight dark matter (either scalar or pseudoscalar). In many cosmological scenarios it is assumed that the ultralight dark matter field $\phi$ (with mass $m$) is coherently oscillating on cosmological scales.  A pseudscalar axion-like (ALP) dark matter field $\phi$ can couple to the $U(1)$ gauge field $A_{\mu}$ of electromagnetism via the interaction Lagrangian
\be
{\cal{L}}_{\rm{int}} \, = \, g_{\phi, \gamma} F \wedge F \, ,
 \ee
where $g_{\phi \gamma}$ is a coupling constant having inverse mass dimensions.  In the absence of plasma effects, the resulting equation of motion for the comoving Fourier modes of $A_{\mu}$ is (once again neglecting the expansion of space) 
\be \label{modeeq}
{\ddot{A}}_{k, \pm} + \bigl( k^2 \pm k g_{\phi \gamma} a(t) {\dot{\phi}} \bigr) A_{k, \pm} \, = \, 0 \, ,
 \ee
where $\pm$ indicates the two different polarization modes. We see immediately \cite{BJF} that, while $\phi$ is oscillating coherently about the minimum of its potential,  there is a tachyonic instability. For modes with $k < k_c$, where
\be \label{kc}
 k_c(\eta) \, = \, g_{\phi \gamma} m \Phi a (t) \, ,
 \ee
the amplitude increases exponentially.  This instability is screened by the plasma before the time of recombination,  but will be operational after that.  Since the instability is exponential, we expect the energy transfer from $\phi$ to $A_{\mu}$ to be rapid on a Hubble time scale.

If the coupling constant $g_{\phi \gamma}$ is not too small relative to its experimental upper bound (a benchmark value being 
\be
g_{\phi \gamma} \, = \, {\tilde{g}}_{\phi \gamma} 10^{-10} {\rm{GeV}}^{-1} 
\ee
with ${\tilde{g}}_{\phi \gamma} \sim 1$), and for a mass $m \sim 10^{-18} {\rm{eV}}$ corresponding to ultralight dark matter, it can be shown \cite{BJF} that this mechanism leads to the generation of magnetic fields on cosmological scales of sufficient amplitude to explain observations (see e.g. \cite{Brev} for review articles on cosmological magnetic fields).  The generation process happens right after the time of recombination. A similar mechanism also works to scalar dark matter \cite{Vahid}. The same mechanism was earlier explored in the context of inflationary magnetogenesis in several references (see e.g. \cite{inflmag}). In this case, $\phi$ is the inflaton field.

There are other interesting examples of oscillating background fields that lead to amplifications via parametric resonance instability,
for example, production of entropy fluctuations during inflationary reheating \cite{Evan}. 
 An oscillating scalar or pseudscalar condensate  introduces small amplitude oscillations in the cosmological scale factor about the background value, which
 can  lead to resonance induced by the dynamics of the scale factor (see e.g. \cite{Mesbah, Ramos2}).  Likewise, oscillating gravitational waves can also lead to  
 enhanced photon production \cite{Chunshan}. 

\section{Conclusions}  
 
We have given an account  from a personal perspective
 of early developments that led to the appreciation that in many models the first stage of energy transfer from the inflaton to matter is driven by a parametric resonance instability, or (p)reheating for short.
 The mathematics involved in preheating turns out to have a wide range of applications in cosmology and beyond and
 some examples  have been mentioned, with an emphasis on research that one of us (RB) has worked on. Around the time of
 the publication of our first reheating paper in 1990, but before it was recognized that reheating was an interesting process,
 the research program of the other one of us (JT) transitioned to an emphasis on black holes; then one day a colleague mentioned that 
 the long ago paper had
 accrued a significant number of citations. All in all, a career thread that could be summarized as ``what a long strange trip it's been"(\cite{GD1970}).

%%%%%%%%%%%%%%%%%%%%%%%%%%%%%%%%%%%%%%%%%%
\begin{acknowledgements}

The research of RB is supported in part by funds from NSERC and from the Canada Research Chair program.  

\end{acknowledgements}
%%%%%%%%%%%%%%%%%%%%%%%%%%%%%%%%%%%%%%%%%

%%%%%%%%%%%%%%%%%%%%%%%%%%%%%%%%%%%%%%%%%%

\end{document}